\documentclass[aps,pra,superscriptaddress,amsmath,amssymb,preprintnumbers,floatfix,showpacs,11pt]{revtex4}
\usepackage{amssymb}
\usepackage{epsfig}

\begin{document}
\title{ Marginal and density
 atomic Wehrl entropies for the Jaynes-Cummings model }
\author{Faisal A. A. El-Orany}
\email{el_orany@yahoo.com} \affiliation{ Department of Mathematics
and Computer Science, Faculty of Science, Suez Canal University
41522, Ismailia, Egypt }

\begin{abstract}
In this paper,  we develop the notion of the marginal and density
atomic Wehrl entropies  for two-level atom interacting with the
single mode field, i.e. Jaynes-Cummings model. For this system we
show that there are relationships between these quantities and
both of the information entropies and the von Neumann entropy.

\end{abstract}
  \maketitle

%%%%%%%%%%%%%%%%%%%%%%%%%%%%
\section{Introduction}
%%%%%%%%%%%%%%%%%%%%%%%%%%%%

 The entanglement represents one of  the most remarkable
feature of quantum mechanics. For an entangled system it is
impossible to factorize its state in a product of independent
states to describe its parts. In the recent years, the
entanglement has been recognized as a resource for
quantum-information processing \cite{prin,tel1,swap}. Various
types of experiments have been performed to the entanglement in
the quantum systems, e.g. long-distance entanglement \cite{peng},
 ion-photon entanglement \cite{volz}, many photons entanglement \cite{zhao},
  etc. For recent review, the reader can consult \cite{rysz}.

Generally, the entanglement in the quantum systems is investigated
by means of the entropy \cite{neum}. There are various definitions
for the entropy including the von Neumann entropy \cite{neum}, the
relative entropy \cite{vdr}, the generalized entropy \cite{basti},
the Renyi entropy \cite{reny}, the linear entropy, and the Wehrl
entropy \cite{wehrl}. The Wehrl entropy has been introduced in
terms of the Glauber coherent states and Husimi $Q$-function. In
the classical limit (, i.e. $\hbar\rightarrow 0$) the von Neumann
entropy tends to the Wehrl entropy \cite{beret}. The Wehrl entropy
 has been successfully applied in the description of
different properties of the quantum optical fields such as
phase-space uncertainty \cite{mira1,{mira2}}, quantum interference
\cite{mira2}, decoherence \cite{deco,{orl}},
 a measure of noise \cite{mira3}, etc. Additionally, it
 has been applied to the dynamical systems, e.g.
the evolution of the radiation field with
 the Kerr-like medium  \cite{jex}
and  with the two-level atom  \cite{orl}, i.e. the Jaynes-Cummings
model (JCM) \cite{jay1}. For the JCM it has been found that the
Wehrl entropy is very sensitive to the phase-space dynamics of
$Q$-function. Also it illustrates the loss of coherence with the
upper limit for the phase randomization during the evolution of
the radiation field \cite{orl}.  The concept of the atomic Wehrl
entropy has been developed \cite{karol} and applied to the JCM
\cite{obad}.  Quite recently, it has been analytically proved that
the linear entropy, the von Neumann entropy and the atomic Wehrl
entropy provide identical information on the entanglement in the
JCM \cite{faiar}. On the other hand, the concept of the phase
density of the Wehrl entropy and/or the Wehrl density distribution
for optical fields has been given in \cite{mira3}. It has been
shown that the Wehrl density distribution clearly describes:
states with random phase, states with a partial phase, phase
locking and phase bifurcation of quantum states of light
\cite{mira3}. Inspired by the concept of the Wehrl density
distribution for the field  we introduce--in the present
paper--the marginal and density atomic Wehrl entropies for the
JCM.  We show that these quantities can reduce to the information
entropies, which are basically used in the treatment of  the
entropy squeezing \cite{fang}. Also they can provide information
on the von Neumann entropy. These are interesting results
motivated by the importance of the JCM in the quantum optics
\cite{jay1}. As is well known that the JCM  can be implemented by
several means, e.g. the one-atom mazer \cite{remp} and the trapped
ion \cite{vogel}.

We perform the study in the following order. In section 2, we
describe the system under consideration and derive the main
relations and equations including the information entropies. In
section 3 we develop the notion of the marginal atomic Wehrl
entropies. In section 4 we give the explicit forms for the density
atomic Wehrl  entropies and discuss their connection with the
information entropies.

%%%%%%%%%%%%%%%%%%%%%%%%%%%%
\section{Model formalism and basic relations}
%%%%%%%%%%%%%%%%%%%%%%%%%%%%
In this section, we give the Hamiltonian model, its wave-function
and the definition of the atomic $Q$-function. Additionally, we
investigate the evolution of the information entropies and the von
Neumann entropy.

Without the loss of generality, we restrict the attention to the
simplest form of the JCM, which is the two-level atom interacting
with the single cavity mode. In the rotating wave  and dipole
approximations the Hamiltonian governing  this system is:

%%%%%%%%%%%%%%%%%%%%%%%%%%%%%%%%%%%%%%%%%%%%%%%%%%%%%%%%%%%%%%%%%%%%%%%%
\begin{eqnarray}
\begin{array}{lr}
\hat{H}=\hat{H}_0+\hat{H}_i\\
\\
 \hat{H}_0 =
\omega_{0}\hat{a}^{\dagger}\hat{a}+
\frac{1}{2}\omega_{a}\hat{\sigma}_{z},\quad \hat{H}_i=
\lambda(\hat{a}\hat{\sigma}_{+} + \hat{a}^{\dagger
}\hat{\sigma}_{-}),
 \label{6}
 \end{array}
\end{eqnarray}
%%%%%%%%%%%%%%%%%%%%%%%%%%%%%%%%%%%%%%%%%%%%%%%%%%%%%%%%%%%%%%%%%%%%%%%%%%%
where $\hat{H}_0\quad (\hat{H}_i)$ is the free (interaction) part,
$\hat{\sigma}_{\pm}$ and $\hat{\sigma}_{z}$ are the Pauli spin
operators; $\omega_{0}$ and $\omega_{a}$ are the frequencies of
the cavity mode and the atomic transition, respectively, $\hat{a}
\quad (\hat{a}^{\dagger})$ is the annihilation (creation) of the
cavity mode, and $\lambda$ is the atom-field coupling constant. In
(\ref{6}) we have set $\hbar=1$ for convenience. We assume that
$\omega_{0}=\omega_{a}$ (, i.e. the resonance case), the field is
initially in the coherent state $|\alpha\rangle$ with real
$\alpha$ and the atom is in the superposition of the excited and
ground atomic states as:

\begin{equation}\label{asup}
|\vartheta\rangle=\cos\vartheta |e\rangle +\sin\vartheta|g\rangle,
\end{equation}
where  $|e\rangle \quad (|g\rangle)$ stands for the excited
(ground) atomic state and $\vartheta$ is a  phase.
 Under these conditions, the dynamical wave function of the
system in the interaction picture can be expressed as:
\begin{equation}
|\Psi (T)\rangle= \sum\limits_{n=0}^{\infty}  \left[G_1(n,T)
|e,n\rangle +G_2(n,T)|g,n+1\rangle\right],
 \label{a8a}
 \end{equation}
 where
\begin{eqnarray}
\begin{array}{lr}
C_n=\frac{\alpha^n}{\sqrt{n!}} \exp(-\frac{1}{2}\alpha^2),
 \qquad T=t\lambda, \\
\\
G_1(n,T)= C_n \cos\vartheta \cos(T\sqrt{n+1})-i
C_{n+1}\sin\vartheta \sin(T\sqrt{n+1}),\\
\\
G_2(n,T)=  C_{n+1}\sin\vartheta \cos(T\sqrt{n+1}) -i C_n
\cos\vartheta \sin(T\sqrt{n+1}). \label{a8an}
 \end{array}
\end{eqnarray}
For reasons will be made clear shortly, we give the expectation
values for the atomic set operators $\{\hat{\sigma}_x,
\hat{\sigma}_y,
 \hat{\sigma}_z\}$ associated with the state (\ref{a8a}) as:
\begin{eqnarray}
\begin{array}{lr}
\langle\hat{\sigma}_{z}(T)\rangle=\sum\limits_{n=0}^{\infty}
\left[|G_1(n,T)|^2-|G_2(n,T)|^2\right],\\
\\
 \langle
\hat{\sigma}_{x}(T)\rangle=2 \textrm{Re}\sum\limits_{n=0}^{\infty}
G^*_1(n+1,T)G_2(n,T),\\
\\
\langle \hat{\sigma}_{y}(T)\rangle=
2\textrm{Im}\sum\limits_{n=0}^{\infty} G^*_1(n+1,T)G_2(n,T),
 \label{ll6y}
 \end{array}
\end{eqnarray}
where $\textrm{Re}$ and $\textrm{Im}$ stand for real and imaginary
parts of the complex quantity. Additionally, the von Neumann
entropy for the JCM can be evaluated as \cite{faiar}:
\begin{eqnarray}
\begin{array}{lr}
\gamma(T) =-\frac{1}{2}[1+\eta(T)]
 {\rm ln}
[\frac{1}{2}+\frac{1}{2}\eta(T)]- \frac{1}{2}[1-\eta(T)]
{\rm ln} [\frac{1}{2}-\frac{1}{2}\eta(T)],\\
\\
\eta(T)= \sqrt{\langle\hat{\sigma}_x(T)\rangle^2+
\langle\hat{\sigma}_y(T)\rangle^2+
\langle\hat{\sigma}_z(T)\rangle^2}.
 \label{law3}
 \end{array}
\end{eqnarray}
As is well known that the von Neumann entropy  is basically used
for quantifying the entanglement, where  $\gamma(T)=0$ for
disentangled and/or pure states and $\gamma(T)=0.693$ for
maximally entangled bipartite, i.e. $0\leq\gamma(T)\leq {\rm
ln}2$.
 We conclude this part by shedding  the light on the
information entropies for two-level system (, i.e $N=2$) described
by the density matrix $\hat{\rho}_a$. The probability distribution
of two possible outcome of measurements of the operator
$\hat{\sigma}_k$ is:
\begin{equation}
P_{j}(\hat{\sigma}_k)=\langle
\psi_{kj}|\hat{\rho}_a|\psi_{kj}\rangle,\qquad j=1,2;\qquad
k=x,y,z,
 \label{en4}
\end{equation}
where $|\psi_{kj}\rangle$ are the eigenstates of
$\hat{\sigma}_k$. In this case the associated information
entropies are:
\begin{equation}
H(\hat{\sigma}_k)=-\sum\limits_{j=1}^{2}P_{j}(\hat{\sigma}_k) {\rm
ln}P_{j}(\hat{\sigma}_k),
 \label{en6}
\end{equation}
where $0\leq H(\hat{\sigma}_k)\leq {\rm ln}2$. It is obvious that
 $H(\hat{\sigma}_k)$ has the same limitations as $\gamma(T)$. It
is worth mentioning that the information entropies are frequently
used in the literatures, e.g., \cite{fang} in the investigation of
the entropy squeezing, in particular, for systems satisfying
 $\langle\hat{\sigma}_z(T)\rangle=0$. For these systems the standard
uncertainty relation of the atomic operators
 fails to provide any useful information on the atomic
system. This difficulty has been overcome  using entropic
uncertainty relation \cite{hirs,{maa}}, which is related to the
information entropies (\ref{en6}). Next, using the short-hand
notations $b =\langle\hat{\sigma}_x(T)\rangle,
c=\langle\hat{\sigma}_y(T)\rangle, h=
\langle\hat{\sigma}_z(T)\rangle$ the relations (\ref{en6}) can be
easily evaluated as:
\begin{eqnarray}
\begin{array}{lr}
H(b)=-\frac{1}{2}(1+b){\rm ln} (\frac{1}{2}+\frac{b}{2})
-\frac{1}{2}(1-b){\rm ln} (\frac{1}{2}-\frac{b}{2}),\\
\\
H(c)=-\frac{1}{2}(1+c){\rm ln} (\frac{1}{2}+\frac{c}{2})
-\frac{1}{2}(1-c){\rm ln} (\frac{1}{2}-\frac{c}{2}),\\
\\
H(h)=-\frac{1}{2}(1+h){\rm ln} (\frac{1}{2}+\frac{h}{2})
-\frac{1}{2}(1-h){\rm ln} (\frac{1}{2}-\frac{h}{2}).
 \label{IaI}
\end{array}
\end{eqnarray}
%%%%%%%%%%%%%%%%%%%%%%%%%%%%%%%%%%%%%%%%%%%%%%%%%%%%%%%%%%%%%%%
\begin{figure}
  \vspace{0cm}
\centerline{\epsfxsize=16cm \epsfbox{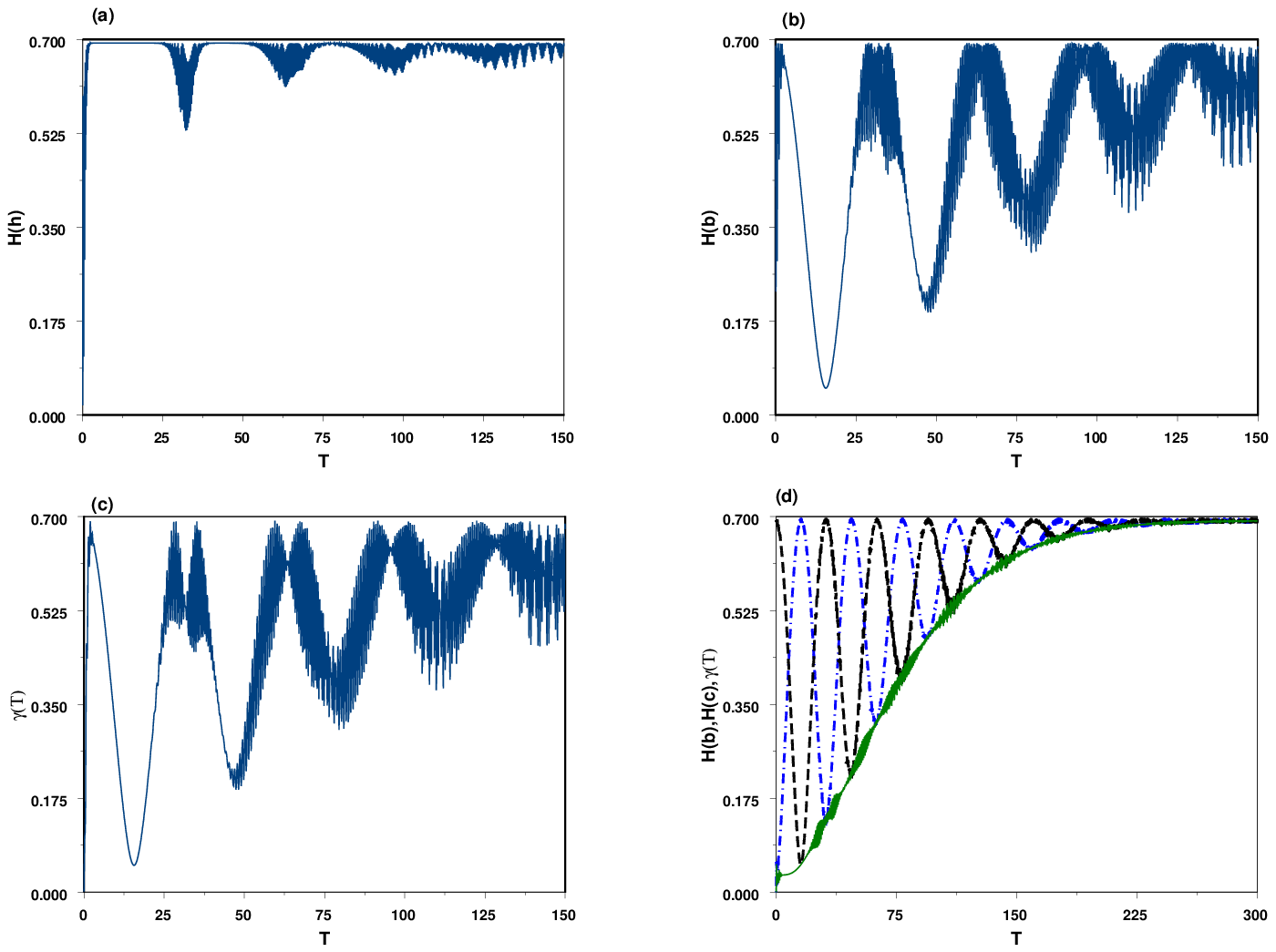} } \vspace{.1cm}
\caption{ Evolution of the information entropies and von Nuemann
entropy  as indicated for $\alpha=5$. Figures (a)--(c) and (d) are
given for $\vartheta=0$ and $\vartheta=\pi/4$, respectively. In
(d)  solid, dashed and dot-dashed curves are given for $\gamma
(T), H(c)$ and  $H(b)$, respectively.}
\end{figure}
%%%%%%%%%%%%%%%%%%%%%%%%%%%%%%%%%%%%%%%%%%%%%%%%%%%%%%%%%%%%
The comparison between expressions (\ref{law3}) and (\ref{IaI})
shows that for particular values of the interaction parameters one
of the information entropies can tend to the von Neumann entropy,
e.g. when $\eta(T)\simeq |\langle \sigma_j(T)\rangle|$.  To see
this and to begin the discussion,  we plot  the von Neumann
entropy and information entropies in Figs. 1 for given values of
the interaction parameters. It is worthwhile mentioning that for
$\vartheta=0, \pi/2$ we have $b=0$ and hence $H(b)={\rm ln}2$. In
this case, the atomic inversion exhibits the revival-collapse
phenomenon (RCP), which is remarkable in Fig. 1(a).  One can
observe that  $H(h)$ provides its maximum value in the course of
the collapse regions. From Fig. 1(b) and (c) one can realize when
the atom is initially in the excited (or ground) state $\gamma(T)$
and $H(c)$ can give quite similar behaviors on the bipartite. The
slight difference between Figs. 1(b) and (c) is that the local
maxima in $H(c)$ are replaced by the local minima in $\gamma(T)$.
Now, the similarity between the behaviors of  $\gamma(T)$ and
$H(c)$ can be explained as follows. When $\alpha$ is real and the
atom is in the excited (or ground) state we always have
$\langle\hat{\sigma}_x(T)\rangle=0$. Additionally, in the course
of the collapse region we have
$\langle\hat{\sigma}_z(T)\rangle=0$, however, during the revival
time the contribution of  $\langle\hat{\sigma}_y(T)\rangle^2$ to
$\eta(T)$ is more effective than that of
$\langle\hat{\sigma}_z(T)\rangle^2$. Thus we can generally
conclude that $\gamma(T)\simeq H(h)$.
 Now, we draw the attention to  Fig. 1(d), which is given for
 $\vartheta=\pi/4$. In this case we have  atomic trapping, i.e.
$\langle\hat{\sigma}_z(T)\rangle\simeq 0$ and hence $H(h)\simeq
{\rm ln}2$.  From Fig. 1(d) one can observe that  $H(b)$ and
$H(c)$ exhibit oscillatory behaviors and  gradually show maximum
values and/or long-living entanglement for large interaction
times. Form the solid curve in  Fig. 1(d) one can observe that
  $\gamma (T)$ is the lower envelope
for   $H(b)$ and $H(c)$, however, for the large interaction times
$\gamma(T)=H(b)=H(c)={\rm ln}2$. This indicates that there is a
systematic loss of coherence for longer interaction times
\cite{orl}. The final remark, the above investigations will be
useful in comparing these quantities with the marginal and density
atomic Wehrl entropies in the next sections.

We close this section by defining the atomic $Q$-function
 $Q_{a}(\theta ,\phi ,T)$ as:

\begin{equation}
Q_{a}(\theta ,\phi ,T)=\frac{1}{2\pi }\left\langle \theta ,\phi
\left\vert \hat{\rho}_{a}(T)\right\vert \theta ,\phi
\right\rangle,  \label{aw1}
\end{equation}%
where
 $\left\vert \theta ,\phi \right\rangle $ is the
atomic coherent state having the form \cite{vie}:
\begin{equation}
\left\vert \theta ,\phi \right\rangle =\cos \left( \theta
/2\right) \left\vert e \right\rangle +\sin \left( \theta /2\right)
\exp (i\phi )\left\vert g \right\rangle \label{aw3}
\end{equation}
with $0\leq \theta\leq \pi, 0\leq \phi \leq 2\pi$. For the
wavefunction (\ref{a8a}) the atomic $Q_a$ function can be
 evaluated  as
\begin{eqnarray}
\begin{array}{lr}
Q_{a}(\theta ,\phi ,T)=\frac{1}{4\pi }[1+\beta(T)],\\
\\
\beta(T) =h\cos \theta +\left[ b\cos \phi +c\sin \phi \right] \sin
\theta. \label{aw4}
\end{array}
\end{eqnarray}
One can easily check that $Q_a$ is normalized. The $Q_a$ can be
interpreted in the following sense. The two different spin
coherent states overlap unless they are directed into two
antipodal points on the Bloch sphere. This is quite different from
that of  $Q$ function of the optical field, which represents the
joint probability distribution for the simultaneous (noisy)
measurements of the two field quadratures \cite{leon}. From
(\ref{aw4}) it is obvious that  $Q_a$ has a complete information
on the set $(b,c,h)$.  In the following sections we use
(\ref{aw4}) to define the marginal and density atomic Wehrl
entropies.

%%%%%%%%%%%%%%%%%%%%%%%%%%%%%%
\section{Marginal atomic Wehrl entropies}
%%%%%%%%%%%%%%%%%%%%%%%%%%%%%%%
In this section we develop the notion of the marginal atomic Wehrl
entropies and show how they can tend to the information entropies
(\ref{IaI}).  In doing so, we  start with  the definitions of the
marginal  atomic $Q_a$ functions as:

\begin{eqnarray}
\begin{array}{lr}
 Q_\theta= \int_{0}^{2\pi} Q_a(\theta,\phi,T) d\phi,  \\
 \\
Q_\phi= \int_{0}^{\pi} Q_a(\theta,\phi,T)
 \sin \theta
 d \theta . \label{mad}
\end{array}
\end{eqnarray}
%%%%%%%%%%%%%%%%%%%%%%%%%%%%%%%%%%%%%%%%%%%%%%%%%%%%%%%%%%%%%%%
\begin{figure}
  \vspace{0cm}
\centerline{\epsfxsize=16cm \epsfbox{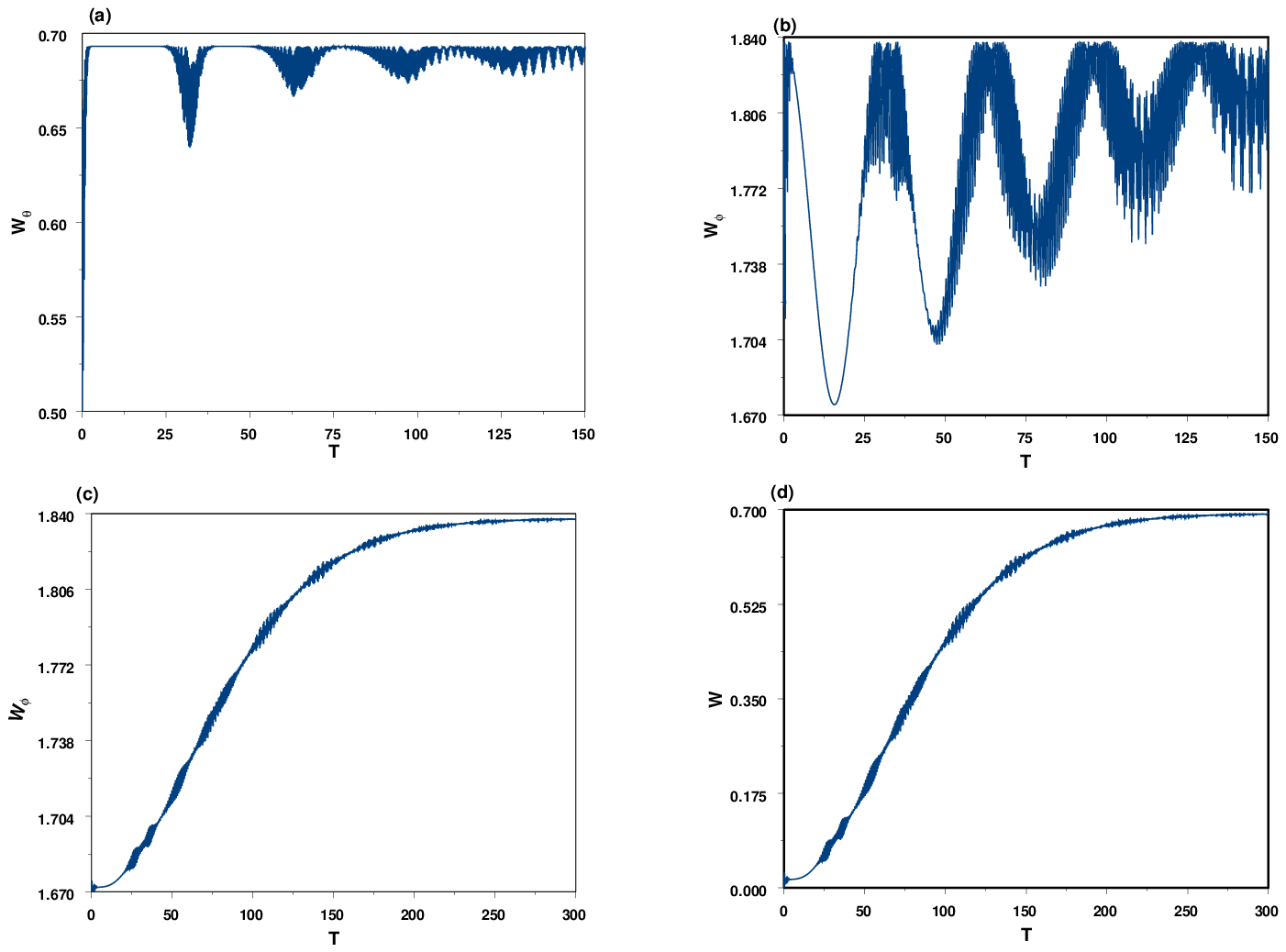} } \vspace{.1cm}
\caption{ Evolution of the marginal atomic Wehrl entropies as
indicated in the figures for $\alpha=5$ against the scaled time
$T$. The figures (a) , (b) and (c)--(d) are given for
$\vartheta=0$ and $\pi/4$, respectively.
 }
\end{figure}
%%%%%%%%%%%%%%%%%%%%%%%%%%%%%%%%%%%%%%%%%%%%%%%%%%%%%%%%%%%%
From (\ref{aw4}) and (\ref{mad}) one can easily obtain:
\begin{eqnarray}
\begin{array}{lr}
 Q_\theta=
 \frac{1}{2}(1+h\cos\theta), \\
\\
Q_\phi= \frac{1}{2\pi}[1+\frac{\pi}{4}(b\cos\phi+c\sin\phi)].
\label{nmad}
\end{array}
\end{eqnarray}
It is obvious that  $ Q_\theta\quad ( Q_\phi)$ includes
information on  $\langle\hat{\sigma}_z(T)\rangle\quad
(\langle\hat{\sigma}_x(T)\rangle,
\langle\hat{\sigma}_y(T)\rangle)$. Now we are in a position to
define the marginal atomic Wehrl entropies as:
\begin{eqnarray}
\begin{array}{lr}
W_{\theta}(T)=-  \int_{0}^{\pi} Q_\theta \ln Q_\theta \sin \theta
d \theta,
 \\\\
W_{\phi}(T)=- \int_{0}^{2\pi} Q_\phi \ln Q_\phi d\phi.
 \label{alwm1}
\end{array}
\end{eqnarray}
As  $W_{\theta}$ and $W_{\phi}$ have been evaluated from $\theta$
and $\phi$ components of  $Q_a$ we call them
 marginal atomic Wehrl entropies. Nevertheless, they  are phase independent.
 It is obvious that
the quantities $W_{\phi}(T)$ and $W_{\theta}(T)$ have the notion
of the entropy, where $Q_{\phi}$ and $Q_\theta$ are always
non-negative quantities (c.f. (\ref{nmad})). In this context,
$W_{\phi}(T)$ and $W_{\theta}(T)$ can be interpreted as being
information measures associated with the components
$\langle\hat{\sigma}_z(T)\rangle$ and
$(\langle\hat{\sigma}_x(T)\rangle,
\langle\hat{\sigma}_y(T)\rangle)$, respectively.
  Substituting (\ref{nmad}) in (\ref{alwm1}) and carrying out
the integration  we obtain:
\begin{eqnarray}
\begin{array}{lr}
W_{\theta}(T)= \ln(2\sqrt{e})+\frac{(1-h)^2}{4h}\ln(1-h)
-\frac{(1+h)^2}{4h}\ln(1+h),\\
\\
=H(h)+ \frac{1}{2}+\frac{(1-h^2)}{4h}{\rm ln}[\frac{1-h}{1+h}],
 \label{alwm2}
\end{array}
\end{eqnarray}
\begin{eqnarray}
\begin{array}{lr}
W_{\phi}(T)=\ln(2\pi)-\sum\limits_{n=0}^{\infty}
\frac{(2n)!}{4^{n+1}[(n+1)!]^2}\xi^{n+1},\\
\\
= {\rm \ln}(2\pi)-\xi
   _3F_2(\{\frac{1}{2},1,1\},\{2,2\};\xi)
 \\
 \\
= {\rm \ln}(2\pi)-1+\sqrt{1-\xi}-{\rm
ln}[\frac{1+\sqrt{1-\xi}}{2}],
 \label{al2}
\end{array}
\end{eqnarray}
where $\xi=\frac{\pi^2(b^2+c^2)}{16}$ and
$_qF_p(\{\tau_1,\tau_2,\cdots
\tau_q\},\{\upsilon_1,\upsilon_2,\cdots \upsilon_p\};\xi)$ is the
generalized hypergeometric function \cite{tabl}.
  In the
derivation of (\ref{al2}) we have used the series expansion of the
logarithmic function  and the following integral identity
\cite{tabl}:

\begin{equation}
\int_{0}^{2\pi}(c_1\sin x +c_2\cos x)^kdx= \left\{
\begin{array}{lr}
0
\;\;&{\rm for}\;k=2m+1 ,\\
2\pi\frac{(2m)!}{4^m (m!)^2} (c_1^2+c_2^2)^m
 \;\;&{\rm
for}\;k=2m,
\end{array}
\right. \label{8a}
\end{equation}
where $c_1, c_2$ are c-numbers and $k$ is a positive integer. The
second and third lines of (\ref{al2}) include  different forms for
the summation in the first line.

From the extreme values of  $h,b,c$ and from the expressions
(\ref{alwm2}), (\ref{al2}) one can obtain the following
inequalities:

\begin{equation}\label{iden2}
\frac{1}{2}\leq W_{\theta}(T) \leq {\rm ln}2,\qquad {\rm
ln}(2\pi)-0.17\leq W_{\phi}(T) \leq {\rm ln}(2\pi).
\end{equation}
The number $0.17$ is value of the series in the first line of
(\ref{al2}), which has been  obtained from its exact form in the
third line. We plot (\ref{alwm2}) and (\ref{al2}) in Figs. 2 for
the given values of the interaction parameters. Comparing figures
(a) and (b) in Figs. 1 with those in Figs. 2 leads to--apart from
the different scales in the Figs. 1 and 2--when the atom is in the
excited (or ground)
 $W_\theta$ and $W_\phi$ can give information on  $H(h)$ and
$H(c)$, respectively.  Nevertheless, when
$\langle\hat{\sigma}_z(T)\rangle\simeq 0$ (, i.e.
$\vartheta=\pi/4$) we have $H(h)=W_\theta={\rm ln}2$, however,
$W_\phi$ gives information on  $\gamma(T)$ (compare the solid
curve in the Fig. 1(d) with the Fig. 2(c)). It is obvious that
$W_\phi$ stabilizes at a certain level after a sufficient long
interaction time. In the language of entanglement, when
$W_{\phi}(T)={\rm ln}(2\pi)-0.17$  [or ${\rm ln}(2\pi)$] the
bipartite is disentangled [or maximally entangled]. Next, we treat
the problem of different scales between the marginal atomic Wehrl
entropies and the information entropies. This  can be raised by
redefining  $W_\theta$ and $W_\phi$ to have the limitations of the
corresponding information entropies, i.e. $0\leq H(.)\leq {\rm ln}
2$. With this in mind and from (\ref{iden2}) we obtain:

\begin{eqnarray}\label{iden3}
\begin{array}{lr}
  \widehat{W}_\theta (T)=\frac{{\rm ln}2}{{\rm ln}(\frac{4}{e})}[
  2W_\theta (T)-1], \\
  W(T) = \frac{{\rm ln}2}{{\rm ln}(2\pi)-0.17}[W_\phi (T)-0.17].
  \end{array}
\end{eqnarray}
We close this section by checking the validity of (\ref{iden3}).
As an example we have plotted the rescaled quantity $W$ in Fig.
2(d). The comparison between Fig. 1(d) and Fig. 2(d) is
instructive and shows that $W(T)\simeq \gamma(T)$.

%%%%%%%%%%%%%%%%%%%%%%%%
\section{Density atomic Wehrl entropies }
%%%%%%%%%%%%%%%%%%%%%%%%
In this section we derive the explicit expressions for  the
density atomic Wehrl entropies, which have been numerically
treated, e.g., \cite{obad} in the static regime. Moreover, we
deduce the connections between these quantities and the
information entropies. The density atomic Wehrl entropies can be
defined as:
\begin{eqnarray}
\begin{array}{lr}
Z_{\theta}(T)=- \int_{0}^{2\pi }Q_a(\theta,\phi,T){\rm
ln}Q_a(\theta,\phi,T) d
\phi ,\\
\\
Z_{\phi}(T)=-\int_{0}^{\pi} Q_a(\theta,\phi,T){\rm
ln}Q_a(\theta,\phi,T) \sin \theta
 d \theta.
 \label{lalw4}
\end{array}
\end{eqnarray}
It is evident that  $Z_\theta, Z_\phi$ are phase dependent and
they have the notion of the entropy. The components $Z_\theta$ and
$Z_\phi$ can be interpreted as being the information measures
associated with the directions  $\theta$ and $\phi$, respectively.
In this respect,  they may also be called geometric information
entropies. Substituting (\ref{aw4}) in (\ref{lalw4}) and carrying
out the integration we obtain the following expressions:

\begin{eqnarray}
\begin{array}{lr}
Z_{\theta}(T)=(1+h\cos\theta)\frac{{\rm ln}(4\pi)}{2}-\frac{1}{2 }
\{h\cos\theta + \sum\limits_{n=2}^{\infty}
\sum\limits_{r=0}^{[\frac{n}{2}]}\frac{(-1)^n(n-2)!}{(n-2r)!(r!)^2
4^r}\\
\\\times(h\cos\theta)^{n-2r}\sin^{2r}\theta (b^2+c^2)^r\},
  \label{alw4}
\end{array}
\end{eqnarray}

\begin{eqnarray}
\begin{array}{lr}
Z_{\phi}(T)=\frac{1}{4\pi}[2+\frac{\pi \varepsilon}{2}]{\rm ln}(4\pi)-\frac{\varepsilon}{8 } \\
\\
+ \sum\limits_{n=1}^{\infty} \sum\limits_{r=0}^{n}
\sum\limits_{s=0}^{n- r}\frac{(2n-1)!(n-r)!(-1)^s h^{2(n-r)}
\varepsilon^{2r+1}}{(2r+1)!(2n-2r)!(n-r-s)!s!(2s+2r+3)4^{s+r+2}
\beta(s+r+2,s+r+2)}\\
\\
-\frac{1}{2\pi}\sum\limits_{n=1}^{\infty} \sum\limits_{r=0}^{n}
\sum\limits_{s=0}^{r}\frac{(2n-2)!r!(-1)^s h^{2(n-r)}
\varepsilon^{2r}}{(2r)!(2n-2r)!(r-s)!s!(2n+2s-2r+1)},
 \label{alw4aa}
\end{array}
\end{eqnarray}
where $\beta(.)$ is the Beta function and $\varepsilon=b\cos\phi+
c\sin\phi$. In the derivation of (\ref{alw4}) and (\ref{alw4aa})
we  have used  procedures similar to those done for (\ref{al2}) as
well as the following identity  \cite{tabl}:
\begin{equation}\label{den4}
    \int_{0}^{\pi} \sin^{m-1}x
    dx=\frac{\pi}{2^{m-1}m\beta(\frac{m+1}{2},\frac{m+1}{2})}.
\end{equation}
From (\ref{alw4}) and (\ref{alw4aa}) one can realize that each of
 $Z_{\theta}$ and $Z_{\phi}$  can  give  information on
 the atomic components, i.e. $h,b,c$. This is in contrast to the
marginal atomic Wehrl entropies (c.f. (\ref{alwm2})-(\ref{al2})).
Also their limitations  are sensitive to the phase as well as the
initial atomic states. We have numerically checked this fact.

Next, we show
 how $Z_{\theta}$ and $Z_{\phi}$ can be connected with the information entropies as well as
$\gamma(T)$. For instance, throughout straightforward calculations
one can easily show:
\begin{equation}\label{ssq}
Z_{\theta=0}(T)+Z_{\theta=\pi}(T)=H(h)+
 {\rm \ln}(2\pi).
\end{equation}

\begin{eqnarray}
\begin{array}{lr}
 Z_{\theta=\pi/2}(T)=\frac{1}{2}{\rm
\ln}(4\pi)-\frac{1}{8}
\sum\limits_{n=0}^{\infty}\frac{(2n)!\bar{\xi}^{n+1}}{4^n[(n+1)!]^2}\\
\\
=\frac{1}{2}{\rm
\ln}(4\pi)-\frac{1}{2}+\frac{1}{2}\sqrt{1-\bar{\xi}}-\frac{1}{2}
{\rm ln}[\frac{1+\sqrt{1-\bar{\xi}}}{2}], \label{ssq}
\end{array}
\end{eqnarray}
where $\bar{\xi}=b^2+c^2$. The series in the first line of
(\ref{ssq}) is similar to that in the (\ref{al2}). Thus
 the comparison between
(\ref{al2}) and (\ref{ssq}) shows that $Z_{\theta=\pi/2}(T)$  can
carry information on the von Neumann entropy. To be more specific,
from (\ref{ssq}) we can obtain the following rescaled density
atomic Werhl entropy:

\begin{equation}\label{iden5}
  \widehat{Z}_{\theta=\pi/2} (T) =\frac{{\rm ln}2}{0.15}[
 Z_{\theta=\pi/2}(T)-\frac{1}{2}{\rm ln}(4\pi)+0.15],
 \end{equation}
where  the number $0.15$ is  obtained form
 (\ref{ssq}) using the extreme values of the $b, c$.
 We have numerically found that
  $\widehat{Z}_{\theta=\pi/2} (T)\simeq \gamma (T)$.
Now, we draw the attention to  $Z_\phi$. When
$\varepsilon\rightarrow 0$ (, i.e. for $b=0$ and $\phi=0$) the
expression (\ref{alw4aa}) reduces to
%%%%%%%%%%%%%%%%%%%%%%%%%%%%%%%%%%%%%%%%%%%%%%%%%%%%%%%%%%%%%%%
\begin{figure}
  \vspace{0cm}
\centerline{\epsfxsize=16cm \epsfbox{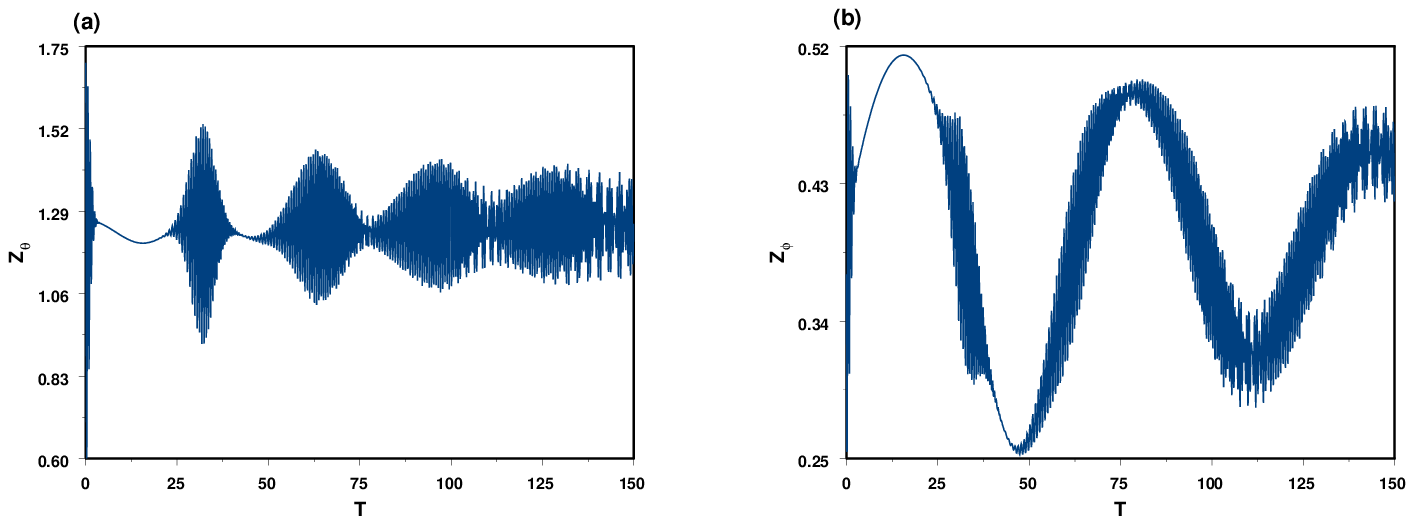} } \vspace{.1cm}
\caption{ Evolution of the density atomic Wehrl entropies   as
indicated in the figures for $(\alpha,\vartheta)=(5,0)$ and
$\theta=\phi=\pi/4$ against the scaled time $T$.
 }
\end{figure}
%%%%%%%%%%%%%%%%%%%%%%%%%%%%%%%%%%%%%%%%%%%%%%%%%%%%%%%%%%%%

\begin{equation}\label{ssqh}
Z_\phi(T)=\frac{1}{2\pi}\{{\rm \ln}(2\pi)+
H(h)+\frac{1}{2}+\frac{(1-h^2)}{4h}{\rm ln}[\frac{1-h}{1+h}]
 \}.
\end{equation}
Also when $h\simeq 0$ (, i.e. the  atomic trapping case) the
expression (\ref{alw4aa}) can give information on  $b$ or $ c$
based on the value of $\phi$.

We close this section by studying numerically the case for which
two or all of the components $(b,c,h)$ give comparable
contribution to the density atomic Wehrl entropies (see Figs. 3).
In these figures we have taken $\theta=\phi=\pi/4, \vartheta=0$.
It is obvious that in the evolution of   $Z_{\theta=\pi/4}\quad
(Z_{\phi=\pi/4}) $ the behavior  of
$\langle\sigma_z(T)\rangle\quad (\langle\sigma_y(T)\rangle)$ is
dominant. It seems that this is related to the leading terms in
the expressions (\ref{alw4}) and (\ref{alw4aa}).

In conclusion, in this article we have developed the notion of the
marginal and density  atomic Wehrl entropies for the JCM. We have
shown that there are relationships between these quantities and
both of the information entropies and von Neumman entropy. The
marginal (density) atomic Wehrl entropies are phase independent
(dependent) and have (do not have) clear limitations. Furthermore,
the marginal (density) atomic Wehrl entropies  can be used as the
information measures associated with the atomic components
(orientations $\theta$ and $\phi$). Finally, we have derived
various analytical relations.

%%%%%%%%%%%%%%%%%%%%%%%%%%%%%%%%%%%%%%%%%%%%%%%%%%%%%%%%%%%%%
\section*{ Acknowledgement}
%%%%%%%%%%%%%%%%%%%%%%%%%%%%%%%%%%%%%%%%%%%%%%%%%%%%%%%%%%%%%%

 I would like to thank the first referee, who I do not know, for his fruitful
 comments. He found an error and corrected. The closed forms
in  (\ref{al2}) and (\ref{ssq}) were given by him.


\begin{thebibliography}{200}



\bibitem{prin} Benenti G, Casati G and Strini G 2005 "Principle of Quantum Computation and
Information" (World Scientific, Singapore).
\bibitem{tel1} Bennet C H,  Brassard G,  Crepeau C,  Jozsa R,
Peresand A and  Wootters W K 1993 {\it Phys. Rev. Lett.} {\bf 70}
1895.

\bibitem{swap}
Gl\"{o}ckl O, Lorenz S, Marquardt C, Heersink J, Brownnutt M,
Silberhorn C, Pan Q, Loock P V, Korolkova N and Leuchs G 2003 {\it
Phys. Rev. A} {\bf 68} 012319; Yang M, Song W and Cao Z-L 2005
{\it Phys. Rev. A } {\bf 71} 034312; Li H-R, Li F-L, Yang Y and
Zhang Q 2005 {\it Phys. Rev. A} {\bf 71} 022314.



\bibitem{peng}  Peng C-Z,  Yang T,  Bao X-H, Zhang J,  Jin X-M,
 Feng F-Y,  Yang B, Yang J,  Yin J,  Zhang Q,  Li N,
 Tian B-L and  Pan J-W
 2005 {\it Phys. Rev. Lett.} {\bf 94} 150501.

\bibitem{volz} Volz U,
Weber M,  Schlenk D,  Rosenfeld W J,  Vrana J,  Saucke K,
Kurtsiefer C, and  Weinfurter H 2006 {\it Phys. Rev. Lett.} {\bf
96} 030404.

\bibitem{zhao} Zhao Z, Chen Y-A, Zhang A-N,  Yang T,  Briegel H and  Pan J-W 2004 {\it
Nature} {\bf 430} 54.

\bibitem{rysz} Horodecki R, Horodecki  P, Horodecki M and
Horodecki K  {\it quant-ph/0702225}.
\bibitem{neum} von Neumann J 1955 "Mathematical Foundations of Quantum
Mechanics" (Princeton University Press, Princeton, NJ).

\bibitem{vdr} Vedral V 2002 {\it Rev. Mod. Phys. } {\bf 74} 197.
\bibitem{basti} Bastiaanns M J 1984 {\it J. Opt. Soc. Am. A} {\bf 1} 711;Tsallis C 1988 {\it J. Stat. Phys.}
{\bf 55} 479.

\bibitem{reny} Renyi A 1970 "Probability Theory" (North Holland, Amsterdam,
1970).

\bibitem{wehrl}Wehrl A 1978 {\it Rev. Mod. Phys.} {\bf 50} 221;
Wehrl A 1991 {\it Rep. Math. Phys.} {\bf 30} 119.

\bibitem{beret} Beretta G P  1984 {\it J. Math. Phys.} {\bf 25} 1507.

\bibitem{mira1} Bu\v{z}ek V, Keitel C H and Knight P L 1995 {\it
Phys. Rev. A} {\bf 51} 2575; Vaccaro J A and Orlowski A 1995 {\it
Phys. Rev. A} {\bf 51} 4172; Watson J B, Keitel C H, Knight P L
and Burnett K 1996 {\it Phys. Rev. A} {\bf 54} 729.
\bibitem{mira2} Bu\v{z}ek V, Keitel C H and Knight P L 1995 {\it
Phys. Rev. A} {\bf 51}  2594.

\bibitem{deco} Anderson A and Halliwell J J 1993 {\it Phys. Rev.
D} {\bf 48} 2753.

\bibitem{orl} Orlowski A, Paul H and Kastelewicz G 1995 {\it Phys.
Rev. A} {\bf 52} 1621.

\bibitem{mira3} Miranowicz A, Matsueda H and
Wahiddin M R B 2000 {\it J. Phys. A: Math. Gen.} {\bf 33} 51519.


\bibitem{jex} Jex I and  Orlowski A 1994 {\it J. Mod. Opt.} {\bf 41}
2301.
\bibitem{jay1} Jaynes E T and Cummings F W
 1963 Proc. IEEE  {\bf 51}  89.


\bibitem{karol} Zyczkowski K 2001  {\it Physica E} {\bf 9} 583.


\bibitem{obad}
 Obada  A-S and  Abdel-Khalek S 2004 {\it J. Phys. A: Math. Gen.} {\bf 37}
  6573; El-Orany F A A, Abdel-Khalek S, Abd-Aty M and Wahiddin M R
  B {\it International J. Theor. Phys. (In
  press); quant-ph/0703043} .


\bibitem{faiar} El-Orany F A A {\it  quant-ph/07042347}.

\bibitem{fang} Fang M-F, Zhou P and Swain S, 2000 {\it J.
Mod. Opt.} {\bf 47} 1043.

\bibitem{remp} Rempe G, Walther H and Klein N 1987 {\it Phys. Rev. Lett.}
{\bf 57} 353.
\bibitem{vogel} Vogel W and De Matos Filho R L 1995 {\it Phys. Rev.
A} {\bf 52} 4214.



\bibitem{hirs} Hirschman I I 1957 {\it Am. J. Math.} {\bf 79} 152;
Bialynickibirula I and Mycielski J 1975 {\it Commun. Math. Phys.}
{\bf 44} 129; Beckner W  1975 {\it Ann. Math.} {\bf 102} 159;
Deutsch D 1983 {\it Phys. Rev. Lett.} {\bf 50} 631.

\bibitem{maa} Maasen H and Unk J B M 1988 {\it Phy. Rev. Lett.} {\bf 60} 1103;
Garret A J M and Gull S F 1990 {\it Phys. Lett. A} {\bf 151} 453;
Sanchez-Ruiz J 1993 {\it Phys. Lett. A} {\bf 173} 233.

\bibitem{vie} Vieira V R and Sacramento P D 1995 Ann. Phys. (N.Y.)
{\bf 242} 188.





\bibitem{leon} Leonhardt U and Paul H 1993 {\it J. Mod. Opt. }
{\bf 40} 1745.

\bibitem{cry1} Ekert A 1991 {\it Phys. Rev. Lett.} {\bf 67} 661;
Cirac J I and  Gisin N 1997 {\it Phys. Lett. A} {\bf 229} 1; Fuchs C
A, Gisin N,  Griffiths R B, Niu   C-S and  Peres A 1997 {\it Phys.
Rev. A} {\bf 56} 1163.
\bibitem{dens} Ye L and Guo G-C 2005 {\it Phy. Rev. A} {\bf 71} 034304;
Mozes S, Oppenheim J and Reznik B 2005 {\it Phys. Rev. A} {\bf 71}
012311.

\bibitem{lamb}  Lambert N,  Emary C and  Brandes T 2004 {\it Phys. Rev.
Lett} {\bf 92} 073602.

\bibitem{vedr} Vedral V 2004 {\it  New. J. Phys.} {\bf 6} 102.

\bibitem{duw} D\"ur W,  Hartmann L,  Hein M,  Lewenstein M and
Briegel H J 2005 {\it Phys. Rev. Lett} {\bf 94} 097203.



\bibitem{shann}
 Shannon C E 1948 {\it Bell Syst. Tech. J. } {\bf 27}
379.



\bibitem{tabl} Gradshteyn S and Ryzhik I M 1994 "Table of Integrals, Series,
and Products" Ed.  Jeffrey A, Fifth edition (Academic Press, Inc.)
 P. 424, 416.



\end{thebibliography}
\end{document}